\documentclass[usenatbib]{mnras}
\usepackage{epsfig}
\usepackage{amsmath,amssymb}
\graphicspath{{figures/}} 
\usepackage{xcolor}
\usepackage{soul}
\usepackage{aas_macros}  
\usepackage{chngcntr}
\graphicspath{{figures/}}

\newcommand{\target}{PHL\,417}
\newcommand{\ltarget}{EPIC\,246373305}
\newcommand{\starget}{SDSS\,J231105.09-013706.0}
\newcommand{\lsiv}{LS\,IV$-14^{\circ}116$}

\newcommand{\kep}{{\em Kepler}}

\newcommand{\ellone}{\ensuremath{\ell}\,=\,1}
\newcommand{\elltwo}{\ensuremath{\ell}\,=\,2}

\newcommand{\uHz}{\,$\mu{\rm Hz}$}

\newcommand{\Msolar}{\mbox{\,$\rm M_{\odot}$}}        

  \newcommand{\Teff}{\mbox{\,\em T$_{\rm eff}$}}         
 \newcommand{\teff}{\mbox{\,$T_{\rm eff}$}}      
\newcommand{\lgcs}{\mbox{\,$\log g / {\rm cm\,s^{-2}}$}}        

\newcommand{\nH}{\mbox{\,$n_{\rm H}$}}                 
  \newcommand{\nHe}{\mbox{\,$n_{\rm He}$}}               
  \newcommand{\kmsec}{\,\mbox{$\mbox{km}\,\mbox{s}^{-1}$}}    
  \def\simge{\mathrel{\raise1.16pt\hbox{$>$}\kern-7.0pt
    \lower3.06pt\hbox{{$\scriptstyle \sim$}}}}           
  \def\simle{\mathrel{\raise1.16pt\hbox{$<$}\kern-7.0pt
    \lower3.06pt\hbox{{$\scriptstyle \sim$}}}}           
\title[PHL\,417]{PHL\,417: a zirconium-rich pulsating hot subdwarf (V366\,Aquarid) discovered in {\it K2} data}
\author[{\O}stensen et al.]
   {R.\,H.\,{\O}stensen,$^{1,2,3}$
    C.\,S.\,Jeffery,$^{4}$\thanks{email: simon.jeffery@armagh.ac.uk}
    H.\,Saio,$^{5}$
    J.\,J.\,Hermes,$^{6}$
    J.\,H.\,Telting,$^{7}$
\newauthor
    M.\,Vu\v{c}kovi\'{c},$^{8}$
    J.\,Vos,$^{9}$
    A.\,S.\,Baran,$^{2}$
and M.\,D.\,Reed$^{3}$\\
$^{1}$Recogito AS, Storgaten 72, N-8200 Fauske, Norway\\
$^{2}$ARDASTELLA Research Group, Institute of Physics, Pedagogical University of Krakow, ul. Podchor\c{a}\.zych 2, 30-084 Krak\'ow, Poland\\
$^{3}$Department of Physics, Astronomy, and Materials Science, Missouri State University, Springfield, MO 65897, USA\\
$^{4}$Armagh Observatory and Planetarium, College Hill, Armagh BT61 9DG, Northern Ireland\\
$^{5}$Astronomical Institute, School of Science, Tohoku University, Sendai 980-8578, Japan\\
$^{6}$Department of Astronomy, Boston University, 725 Commonwealth Ave., Boston, MA 02215, USA\\
$^{7}$Nordic Optical Telescope, Rambla Jos\'e Ana Fern\'andez P\'erez 7, 38711 Bre\~na Baja, Spain\\
$^{8}$Departamento de F\'{i}sica y Astronom\'{i}a, Universidad de Valpara\'{i}so, Avenida Gran Breta\~{n}a 1111, Valpara\'{i}so 2360102, Chile \\
$^{9}$Institut f\"ur Physik und Astronomie, Universit\"at Potsdam, Karl-Liebknecht-Str. 24/25, 14476, Golm, German
}

\date{Accepted 2020 October 5. 
      Received 2020 September 16; in original form 2020 June 3}

\pagerange{\pageref{firstpage}--\pageref{lastpage}}
\pubyear{2020}

\begin{document} 

\maketitle
\label{firstpage}

\begin{abstract}
The {\em Kepler} spacecraft observed the hot subdwarf star \target\ during its extended {\em K2} mission, and the high-precision photometric lightcurve reveals the presence of 17 pulsation modes with periods between 38 and 105 minutes. 
From follow-up ground-based spectroscopy we find that the object has a relatively high temperature of 35\,600\,K, a surface gravity of $\log g / {\rm cm\,s^{-2}}\,=\,5.75$ and a super-solar helium abundance. 
Remarkably, it also shows strong zirconium lines corresponding to an apparent +3.9 dex overabundance compared with the Sun. These properties clearly identify this object as the third member of the rare group of pulsating heavy-metal stars, the V366-Aquarii pulsators.
These stars are intriguing in that the pulsations are inconsistent with the standard models for pulsations in hot subdwarfs, which predicts that they should display short-period pulsations rather than the observed longer periods.
We perform a stability analysis of the pulsation modes based on data from two campaigns with {\em K2}.
The highest amplitude mode is found to be stable with a period drift, $\dot{P}$, of less than $1.1\cdot10^{-9}$ s/s.
This result rules out pulsations driven during the rapid stages of helium flash ignition.
\end{abstract}


\section{Introduction}

Hot subdwarf stars are the evolved remnants of stars that have passed beyond the main-sequence and red-giant-branch stages of evolution, and through one of several possible mass-loss mechanisms they have lost most of their hydrogen envelopes. 
They are found in significant numbers between the main sequence and the white-dwarf cooling track in the Hertzsprung-Russell diagram. 
The bulk of their population is made up of the extreme-horizontal-branch (EHB) stars, whose progenitors ignited helium through the core-helium flash, but other stars also reach the hot-subdwarf population either through interaction with a companion or through a merger of two low-mass white dwarfs \citep[see][for reviews]{heber09,heber16}.

B-type hot subdwarf (sdB) stars pulsate with short (p-mode: 2\,--\,10\,m) and long (g-modes: 30\,--\,120\,m) periods.
Driving is  due to an opacity bump ($\kappa$-mechanism) associated with a local overabundance of iron-group elements at temperatures around $2\times10^5$\,K which results from competition between radiative levitation and gravitational settling  \citep{charpinet97,fontaine03,jeffery06b}. 
The hotter, short-period pulsators are known as V361-Hya stars after the prototype \citep{kilkenny97}, and the cooler long-period pulsators as V1093-Her stars \citep{green03}. 
Pulsations have also been detected in the much hotter sdO stars, but the only pulsators of this type found in the field are the prototype, V499\,Ser \citep{woudt06}, and EO\,Ceti \citep{koen98,ostensen12b}. These helium-poor sdO stars are thought to be post-EHB stars, since core-helium burning sdB stars are expected to move to hotter temperatures as they start to run out of helium in their core, and enter a shell-helium burning phase. 
The pulsations in these stars is not considered to be problematic, since the same $\kappa$-mechanism is expected to operate in this stage, as long as radiative levitation is allowed to work and establish an iron-group-element opacity bump in the driving region \citep{fontaine08}.
As shell-helium burning ceases, post-EHB stars contract and move towards the white-dwarf cooling curve, with increasingly short natural frequencies. 
One sdO star with extremely short pulsation periods ($\sim$30\,s) may be the first example \citep{kilkenny17}, and could represent the first detected post-EHB DAO pulsator, predicted to be excited by the $\epsilon$-mechanism associated with H-shell burning by \citet{charpinet97b}.

For a long time the most mysterious hot-subdwarf variable has been the long-period intermediate helium sdB star \lsiv\ = V366\,Aqr, which was discovered to be a long-period pulsator by \citet{ahmad05}.  
Whilst subdwarf B stars have predominantly hydrogen-rich surfaces, sdO stars show surfaces ranging from hydrogen-rich to extremely hydrogen-poor. 
The latter are often classified He-sdO. 
\citet{naslim10} identified a group of subdwarfs having intermediate surface-helium enrichment, i.e.~between 10\% and 90\% helium by number. 
Most lie close to the sdO/sdB boundary and have thus been variously referred to as intermediate helium subdwarfs, iHe-sdOs and iHe-sdBs.  

\citet{ahmad05} and \citet{green11} recognised that the periods observed in \lsiv\ were incompatible with the same (Fe-group) $\kappa$-mechanism responsible for pulsations in V361-Hya and V1093-Her pulsators. 
A spectroscopic study by \citet{naslim11} revealed that it was not only somewhat enriched in helium, but found it to have a number of unusual absorption lines from germanium, strontium, yttrium and zirconium in its optical spectrum, detected at abundances as high as 10\,000 times solar. 
Other examples of such {\it heavy-metal} stars were later identified \citep{naslim13}. 
\citet{bertolami11} proposed that an $\epsilon$-mechanism present during the core-helium flash ignition phase, rather than the normal $\kappa$-mechanism could be responsible for driving the pulsations in V366\,Aqr \citep[see also][]{battich18}. Recently, \citet{bertolami20} have proposed stochastic excitation by convection generated by the same rapid core-helium flashes as an alternative. Both driving mechanisms require the star to be in the pre-EHB phase.

A second object with pulsations similar to V366\,Aqr was discovered by \citet{latour19b}; Feige\,46. This star has a similar atmosphere to the prototype \citep{latour19a}. Its pulsation periods (2\,200 -- 3\,400\,s) overlap the range found for V366\,Aqr  \citep[][1\,950 -- 5\,100\,s]{green11}. 
Subsequently, \citet{saio19} showed that the $\kappa$-mechanism operating through carbon and oxygen opacities at $\sim1$ million K could account for the  pulsations in both \lsiv\ and Feige\,46, but only if carbon and/or oxygen abundances are substantially enhanced at these temperatures. 

In this paper we present the third object of the V366-Aqr class of pulsators, \target, and the first such object observed with space-based photometry.

\begin{figure}
\epsfig{file=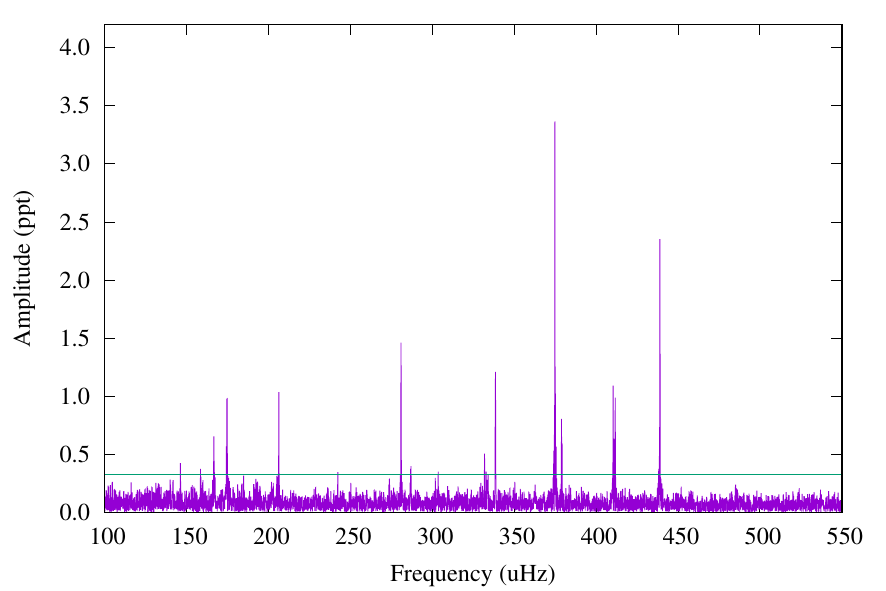,width=1.0\linewidth}\\
\epsfig{file=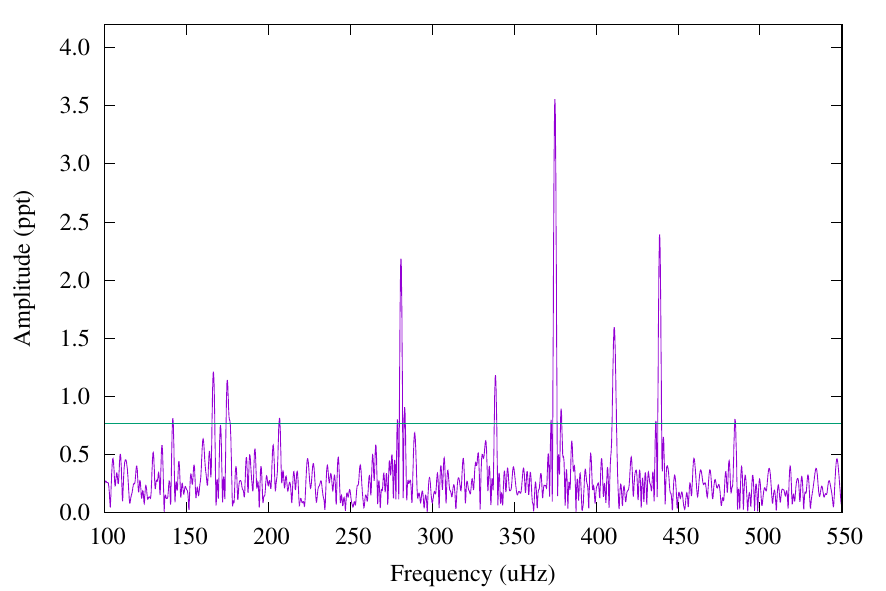,width=1.0\linewidth}
\caption{Amplitude spectra of the K2 datasets of \target\ up to 550\,\uHz\ for C12 (upper panel), and C19 (lower panel). The green line indicates the detection threshold (see text for details).}
\label{fig:kepft}
\end{figure}

\section{K2 observations}

The target of the current paper was first catalogued as \target\ from the survey of \citet{PHLCat}, where it is identified as a faint blue star with a photographic magnitude $B$\,=\,16.9, and colours $B-V$\,=\,--0.2, $U-B$\,=\,--0.4.
It appears in the SDSS catalog \citep{SDSS} as \starget\ with {\em ugriz}\,=\,16.22, 16.52, 17.01, 17.37, 17.68.
It was based on these colours that it was selected as a candidate WD pulsator and proposed as a short-cadence target for the {\em K2} mission, 
where it was designated \ltarget\ in the Ecliptic Plane Input Catalog \citep{huber16}.
The Gaia DR2 catalogue has this star with $G$\,=\,16.792(2) and a parallax of 0.48(9) mas \citep{gaia18.dr2}.

\begin{figure}
\epsfig{file=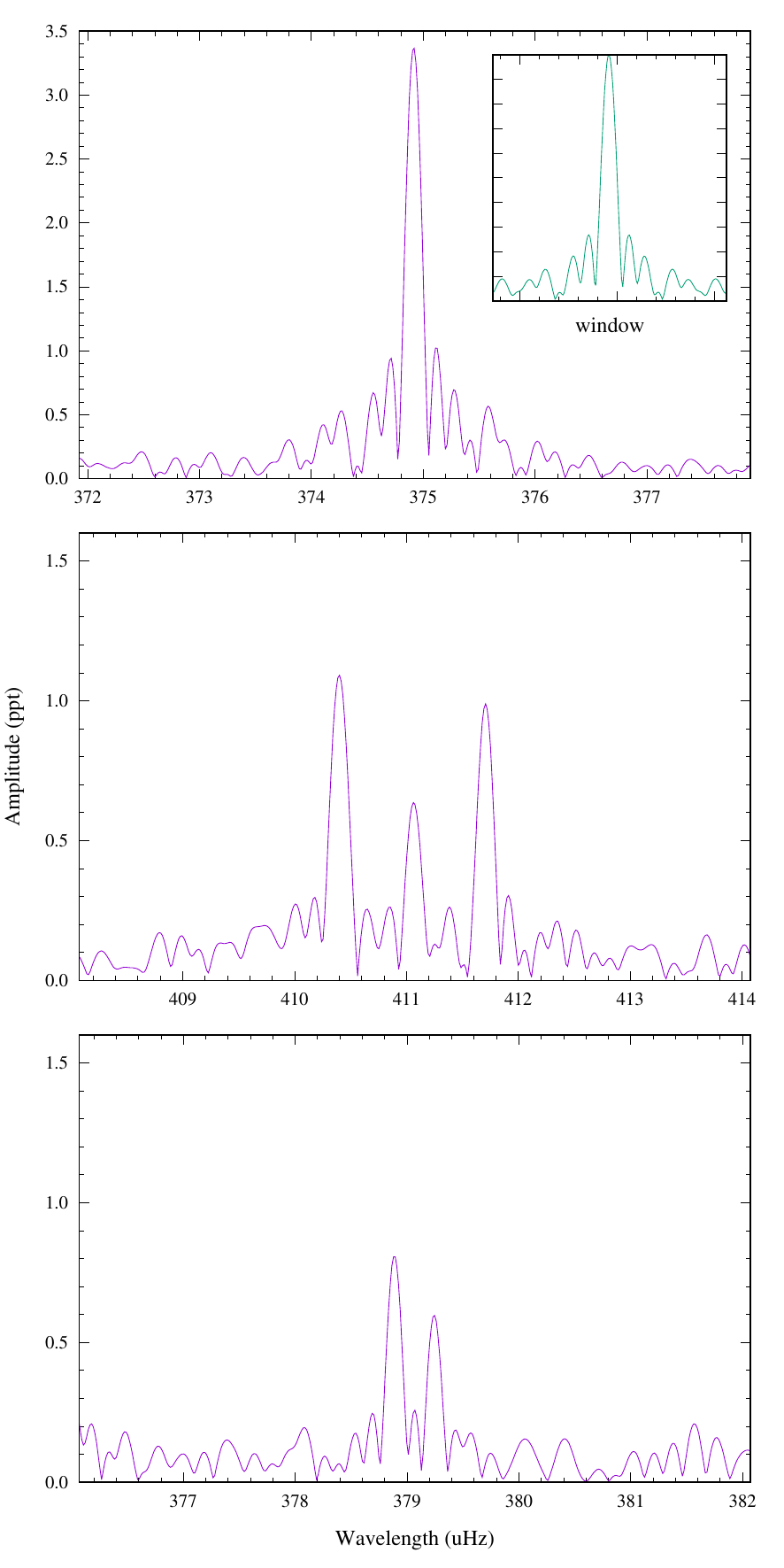,width=1.0\linewidth}
\caption{Expanded sections of the FT of the C12 lightcurve of \target: the largest-amplitude peak ($f_{14}$; top), a triplet ($f_{16}$; middle), and a doublet ($f_{15}$; bottom). The window function is shown inset in the top panel. }
\label{fig:peaks}
\end{figure}

\target\ was observed in short cadence (SC) during Campaigns 12 and 19 of the {\em K2} mission. 
C12 observations started on 2016 Dec 15 and ended on 2017 Mar 4, with a 5-day gap in the 79\,d run. 
The gap, which occurred around 2/3 of the way through the campaign, was not large enough to produce significant multi-peak structure in the window function (Fig.~\ref{fig:peaks}, inset).
C19 ran from 2018 Sep 8 to 23, when the spacecraft ran out of propellant.

For C12,  target pixel files were downloaded from the ``Barbara A. Mikulski Archive for Space Telescopes'' (MAST)\footnote{archive.stsci.edu}, since no pipeline processed SC light curves were available at the time.
The pixel files were processed by first using standard IRAF tasks to extract fluxes, and then using custom Python scripts to decorrelate fluxes in $X$ and $Y$ directions around a 2D mask for the target to remove the light-curve effects caused by the spacecraft’s thruster firings. 
For C19, the pipeline-processed SC light curve was used. 
This lightcurve covers only 7.25\,days of the C19 campaign. 
During the first part of the campaign, the pointing of the telescope was off, and during the last part, the pointing was too erratic to recover useful data. 
The usable data are sufficient to verify the frequencies found in C12.

The Fourier transform (FT) of the resulting light curve is shown in Fig.~\ref{fig:kepft}. 
A number of significant peaks are clearly seen in the low-frequency region of the FT. 
By corollary with analyses of light curves of other hot subdwarfs, where pulsations were established independently from light and velocity variations, and from theoretical models, the multi-peaked FT of the {\it K2} lightcurve of \target\ is strongly suggestive of the multi-period non-radial oscillations seen in V1093-Her variables. 

The RMS $\sigma$ in the featureless frequency region between 1000 and 2000\,\uHz\ was measured to be 0.072 parts per thousand (ppt) and 0.192\,ppt for C12 and C19 respectively. 
This indicates the noise level and is representative for the region where  significant frequencies are found. 
The noise increases slightly at the lowest frequencies. 
The light curves have $n$\,=\,104\,542 (C12) and 10\,466 (C19) good data points. 
For a normal distribution, the 1-in-$n$ chance of  a spurious detection $\sigma$ should be 4.43 (C12) and 3.9 (C19). 
Detection thresholds of 4.5 and 4.0$\sigma$ were adopted for the two datasets, i.e. 0.33 and 0.77\,ppt, respectively.

Taking into account the difference in sampling and resolution, the two FTs are almost identical. The only difference is a peak at 484.8\,\uHz\ which is significant in C19 and only just detectable in C12. It is evident from Fig.~\ref{fig:kepft} that the pulsator is very stable over the two runs. When taking into account that the apertures are somewhat different, and that the background flux may be different, the amplitudes of the main peaks can easily be the same within the errors.

\begin{table}
\caption{\label{t:freqs}\small
Peaks detected in the Fourier transform of the K2 lightcurve of \target.}
\centering
\renewcommand{\arraystretch}{1.2}
\begin{tabular}{llcc}
\hline\hline
ID  & Frequency & Period & Amplitude \\
    & [\uHz] & [s] & [ppt] \\ \hline
$f_1$         & 146.291(11) & 6835.7 & 0.42(6) \\
$f_2$         & 158.589(13) & 6305.6 & 0.43(-) \\
$f_3$         & 166.747(7)  & 5997.1 & 0.66(6) \\
$f_4$         & 174.750(5)  & 5722.5 & 0.97(6) \\
$f_5$         & 206.418(5)  & 4844.6 & 1.04(6) \\
$f_6$         & 242.410(14) & 4125.2 & 0.35(6) \\
$f_7$         & 280.973(3)  & 3559.0 & 1.47(6) \\
$f_8$         & 286.990(-)  & 3484.4 & 0.40(-) \\
$f_9$         & 303.630(13) & 3293.5 & 0.35(6) \\
$f_{10}$      & 331.960(9)  & 3012.4 & 0.50(6) \\
$f_{11}$      & 332.939(13) & 3003.6 & 0.37(6) \\
$f_{12}$      & 338.431(7)  & 2954.8 & 0.74(6) \\
$f_{13}$      & 338.606(4)  & 2953.3 & 1.25(6) \\
$f_{14}$      & 374.914(1)  & 2667.3 & 3.39(6) \\
$f_{15}^{\,-}$& 378.896(7)  & 2639.3 & 0.74(6) \\
$f_{15}^{\,+}$& 379.233(10) & 2636.9 & 0.53(6) \\
$f_{16}^{\,-}$& 410.403(4)  & 2436.7 & 1.12(6) \\
$f_{16}^{\,0}$& 411.064(7)  & 2432.8 & 0.66(6) \\
$f_{16}^{\,+}$& 411.702(5)  & 2428.9 & 0.90(6) \\
$f_{17}$      & 438.941(2)  & 2278.2 & 2.35(6) \\
\hline
\end{tabular}
\end{table}

We summarise the result of the frequency analysis in Table\,\ref{t:freqs}.
Two of the peaks in the FT show  multiple components. 
If due to non-radial oscillations, the multiplets could be consistent with sectoral modes having \ellone\ and 2. 
The triplet ($f_{16}$) around 411\,\uHz\ shows a symmetric splitting of 0.66\,\uHz. 
The doublet ($f_{15}$) at 379\,\uHz\ shows a split of 0.34\,\uHz. 
As this is the narrowest splitting, it most likely corresponds to an \ellone\ g-mode, which would have a normal Ledoux constant $C_{nl}= 0.5$, and corresponding to a rotation period of $\sim$17\,days.
Under conservative assumptions for the radius, this corresponds to an equatorial velocity of $\sim$0.5\,km/s, so
\target\ appears to be a slow rotator,  just like \lsiv\ and Feige\,46 \citep{naslim11,latour19b}.
Without detecting more multiplets, direct mode identification is not possible from rotational frequency splitting.

The period spacing between modes of the same $\ell$ and consecutive radial order has been shown to follow the expected asymptotic sequences with even period spacing of $\sim$250\,s in almost all V1093-Her stars observed with \kep.
Although the main mode and the triplet show a spacing of 234\,s, the sequence is not found to continue, and no other even spacing can be found.

The only sdB pulsators in the temperature region above 35\,kK with high-quality space data are KIC\,10139564, observed during the main {\em Kepler} mission \citep{baran12}, and the {\em K2} target PG\,1315--123 \citep{reed19}.
While both these stars are predominantly p-mode pulsators, they also show a fair number of pulsation periods in the range 1500 -- 4000\,s, which overlaps with the period range found in \target. 
Unlike the periods of the cooler V1093-Her stars, which form evenly spaced sequences as predicted by theoretical models, the periods in KIC\,10139564 and PG\,1315--123 do not follow any clearly evenly spaced sequences, just as we see for \target.

Of particular interest is the phase stability of the pulsation modes, since modes driven by the $\epsilon$-mechanism have been predicted to be changing at a rate of $\dot{P}$\,$\approx$\,$10^{-7}$\,--\,$10^{-4}$ \citep{battich18}.
We can divide the {\em K2} datasets into chunks roughly one week long each, and measure how frequency and phase change in time.
Not one of the modes that have significant amplitudes in the individual chunks shows any sign of a clear drift.
This becomes even more evident when including the C19 data to the picture.
For all modes detected in C19, the C19 period is found to be well within the spread of the measurements from the C12 chunks.
The modes that give the clearest constraint are the highest-amplitude modes.
For $f_{14}$ we see that any change must be slower than $1.1\cdot10^{-9}$ s/s, and for $f_{17}$ we get a limit of $3.0\cdot10^{-9}$ s/s. These numbers are calculated using 3-$\sigma$ uncertainties from a quadratic fit to the observed phase evolution (O-C diagram) for $f_{14}$ and $f_{17}$.

\begin{figure}
  \epsfig{file=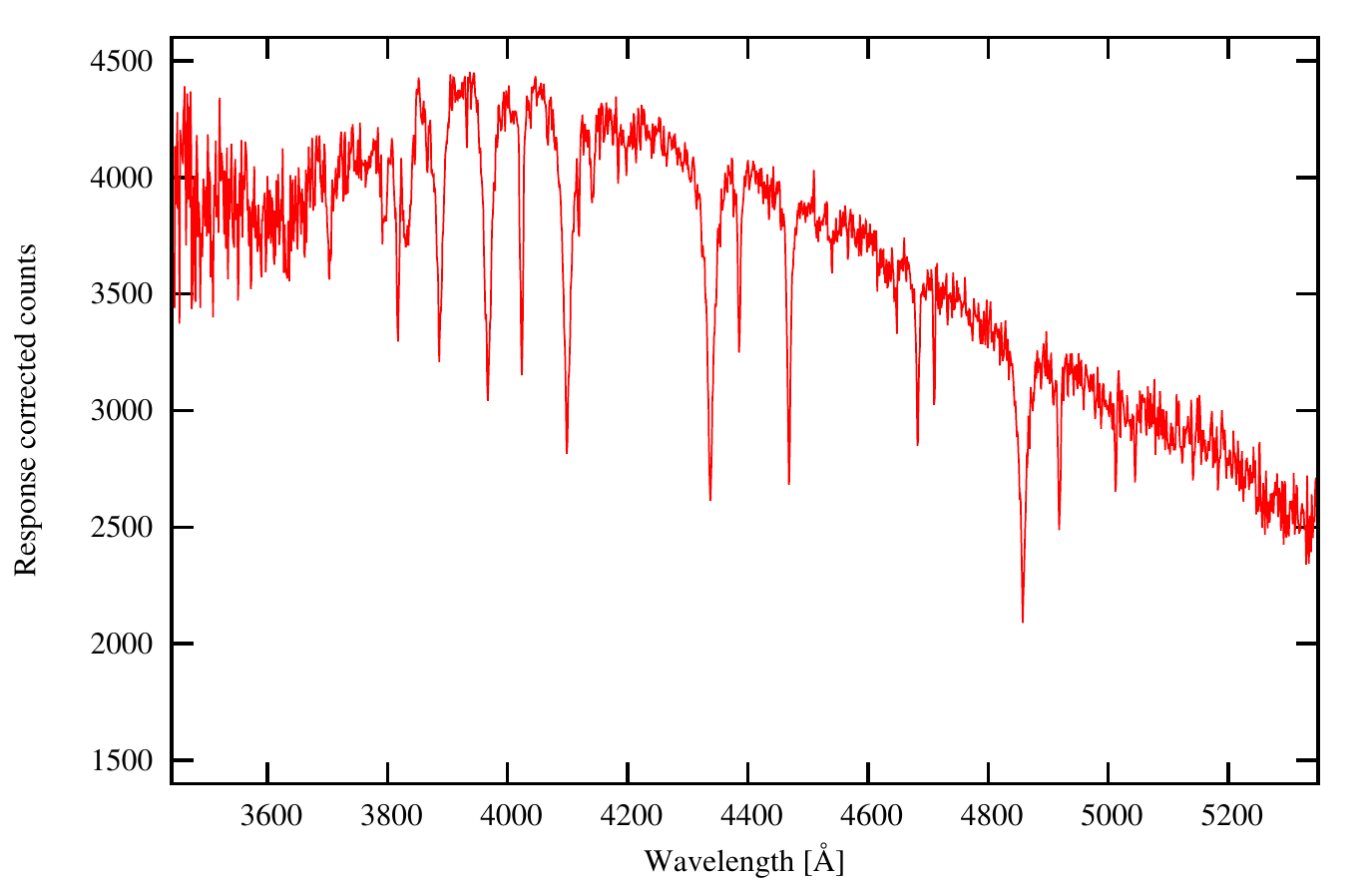,width=1.0\linewidth}
\caption{Sum of the 8 NOT/ALFOSC spectra.}
\label{fig:spek}
\end{figure}

\section{Spectroscopy}

Following notification of sdB-like pulsations, \target\ was added to the spectroscopic \kep\ sdB follow-up survey \citep{telting12b}.
The survey uses the ALFOSC spectrograph on the 2.6-m Nordic Optical Telescope (NOT), with grism \#18 and an 0.5-arcsec slit, giving a resolution R\,=\,2000, which corresponds to 2.2\,\AA\ FWHM.
Eight spectra of \target\ were obtained between 2017 Aug 3 and 2017 Oct 11 (Table\,\ref{t:obs}),
and are more than sufficient to identify any orbital acceleration. 
The data are consistent with no radial velocity variations, within the time-scale covered by the observations, which is too short to rule out long-period binarity with time-scales on the order of one hundred days or more.
The median radial velocity error of the data set is 3.5\,km/s, when using the sharper He lines rather than the Balmer lines.
After applying heliocentric corrections, the spectra were co-added to yield a mean spectrum with a S/N peaking at $\approx83$ at 4000\,\AA\ (Fig.~\ref{fig:spek}).
The radial velocity of the mean spectrum, as measured from He{\sc i} lines, is $-63\pm5$\kmsec. A systematic zero-point error of 10 \kmsec\ is appropriate for ALFOSC. 

The NOT/ALFOSC spectrum of \target\ does not show the typical V1093-Her spectrum expected from initial analysis of the light curve. 
On the contrary, both neutral and ionized helium are strong. 
A preliminary analysis of the first spectrum showed effective temperature (\teff), surface gravity ($g$) and surface helium abundance ($y = \nHe/\nH$) all too high for \target\ to be a V1093-Her variable,
and indicated that the spectrum is similar to that of the helium sdB pulsators \lsiv\ \citep{naslim11} and Feige\,46 \citep{latour19b}.
This  prompted follow-up spectroscopy using the Magellan Echellette \citep[MagE,][]{marshall08}  on the 6.5-m Baade Magellan Telescope on Las Campanas Observatory, Chile. Three spectra with exposures of 1200\,s each were obtained on the night of 18 September 2017 with a 1.0 arcsecond slit yielding a resolution of R \,=\, 4100, see Table\,\ref{t:obs}. In addition to the science spectra, the internal quartz and dome flat field lamps were obtained for pixel response calibration and ThAr lamps spectra were obtained for wavelength calibration. The MagE spectra were reduced using the MASE pipeline \citep{bochanski09} following standard procedures for order tracing, flat field correction, wavelength calibration, optimal source extraction. The extracted spectrum (each order) was divided by the blaze function, which is obtained from the flat field frames. Given the differences in the continuum of the observed star and the blaze correcting function (obtained for the flat) the deblazed orders will inevitably have a certain slope (smooth variation) that will not reflect the shape of the continuum of the observed star. Therefore, we have continuum normalized each order by fitting a low order polynomial to the deblazed flux including an iterative procedure that excludes absorption lines from the fit. Once the orders were continuum normalized they were  co-added in a single spectrum. For the overlapping edges of sequential orders, the average was computed. 
The reduced spectrum shown in Fig.\ref{f:echelle} is the sum of 3$\times$1200\,s, exposures and has a S/N $\approx100$ at 4750\,\AA. 

The wavelength coverage of MagE spans 3080\,--\,8280\,\AA. 

\begin{table}
\caption{\label{t:obs}\small
Log of spectroscopic observations.}
\centering
\renewcommand{\arraystretch}{1.0}
\begin{tabular}{lccc}
\hline\hline
Date-Time  & Exp & S/N & Telescope/Inst. \\
(UT start)    & [s] &      &  \\ \hline
2017-08-12 04:38:21 &  900 & 23 & NOT/ALFOSC \\
2017-08-13 01:58:00 &  900 & 34 & NOT/ALFOSC \\
2017-08-14 01:02:20 &  900 & 31 & NOT/ALFOSC \\
2017-08-20 05:06:49 &  900 & 26 & NOT/ALFOSC \\
2017-09-18 05:19:52 & 1200 & 64 & Magellan/MagE \\
2017-09-18 05:40:16 & 1200 & 64 & Magellan/MagE \\
2017-09-18 06:00:39 & 1200 & 60 & Magellan/MagE \\
2017-09-19 23:40:40 &  900 & 42 & NOT/ALFOSC \\
2017-09-30 22:03:55 &  900 & 27 & NOT/ALFOSC \\
2017-10-11 00:56:59 &  900 & 29 & NOT/ALFOSC \\
2017-10-11 01:31:31 &  900 & 34 & NOT/ALFOSC \\
\hline
\end{tabular}
\end{table}

\begin{table}
\caption[Atmospheric parameters]
   {Atmospheric parameters for \target\ obtained from spectra observed with ALFOSC and MagE. Parentheses show statistical errors in the last digits(s).  Systematic errors due to observation can be estimated by comparing the two sets of measurements.
   }
\label{t:atmos}
\small
\begin{center}
\begin{tabular}{l lll}
\hline
Solution      & ALFOSC &  MagE  \\
\hline
\teff/kK     &  $35.35(6)$ & $35.92(4)$ \\
$\lgcs$      & $5.73(1)$  &  $5.74(1)$  \\
$n_{\rm He}$ &  $0.232(3)$ & $0.259(3)$ \\
$\log y$     &  $-0.520(2)$ & $-0.457(2)$ \\ 
\hline
\end{tabular}
\end{center}
\end{table}

\begin{figure*}
\epsfig{file=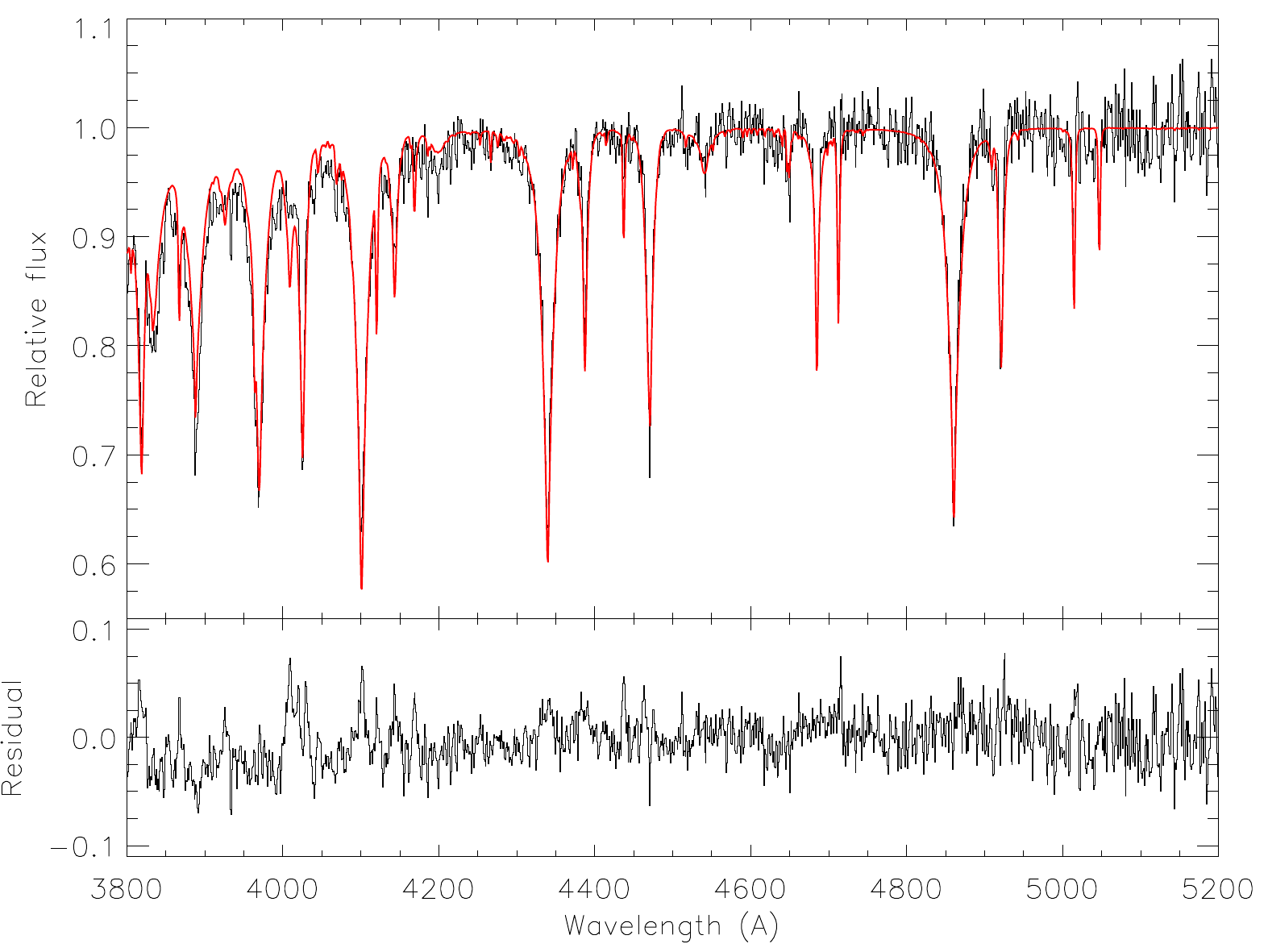,width=0.9\textwidth}
\caption{The ALFOSC spectrum of \target\ (black histogram: upper panel) 
and the best fit (interpolated) model (red, grey in print) having  $\teff=35\,350$\,K, $\lgcs=5.73$,  and $n_{\rm He}=0.23$, $v_{\rm turb}=5\kmsec$.  
The model has been degraded to the instrumental resolution. The residual (observed -- model) is plotted at the same scale in the lower panel.}
\label{f:fit_alfosc}
\end{figure*}

\begin{figure*}
\epsfig{file=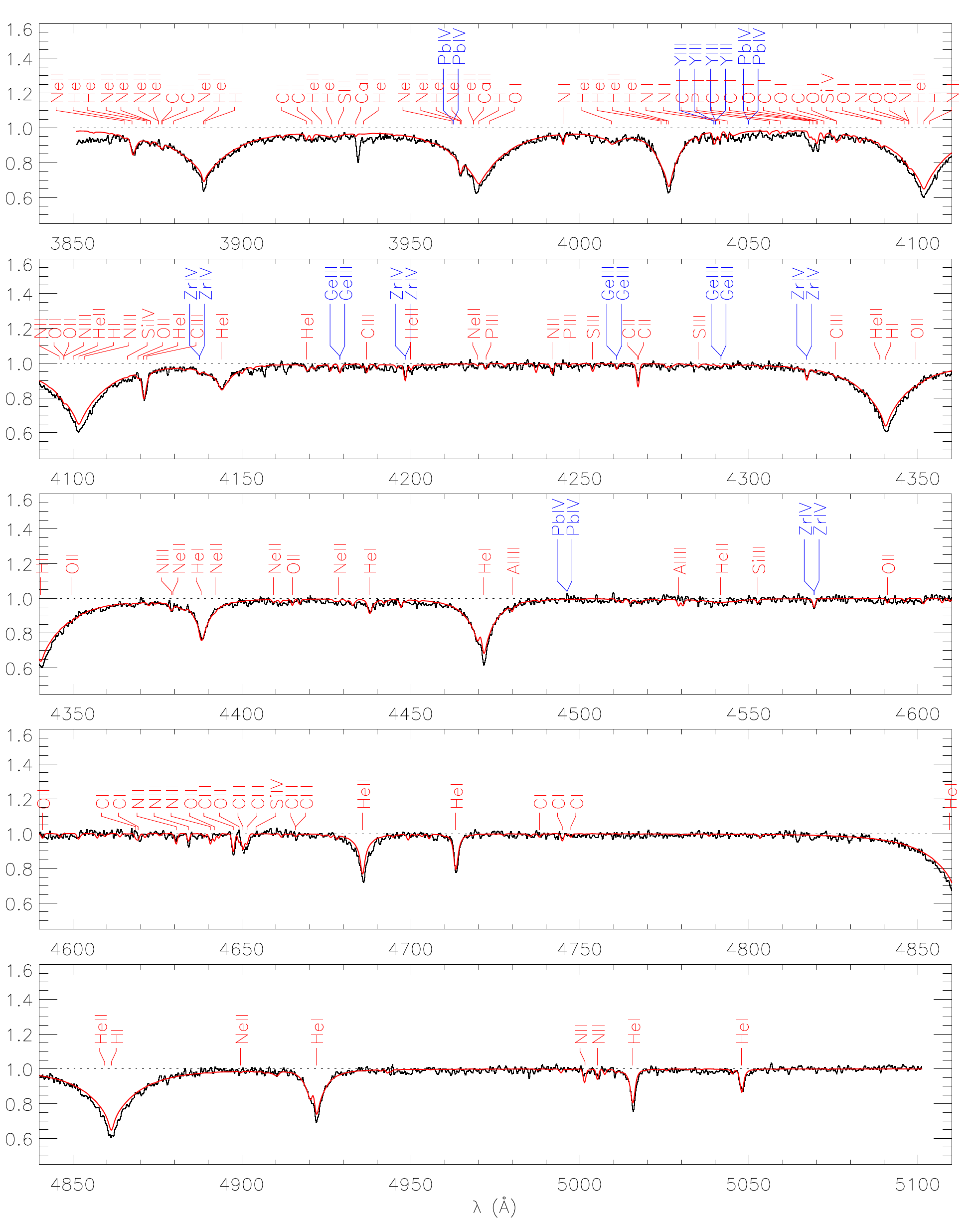,width=0.9\textwidth}
\caption{The MagE spectrum of \target\ (black histogram) and a model having   $\teff=35\,600$\,K, $\lgcs=5.750$, $\nHe=0.23$, $v_{\rm turb}=5\kmsec$  and abundances as given in Table\,\ref{t:abunds}.     
The observed spectrum and model have been resampled at a dispersion of 0.15\AA \,pixel$^{-1}$.
Lines with predicted equivalent widths $>15$\,m\AA\ are labelled. 
The model is convolved with an instrumental profile of 0.8\AA\ (FWHM).}
\label{f:fit_mage}
\end{figure*}

\begin{figure*}
\epsfig{file=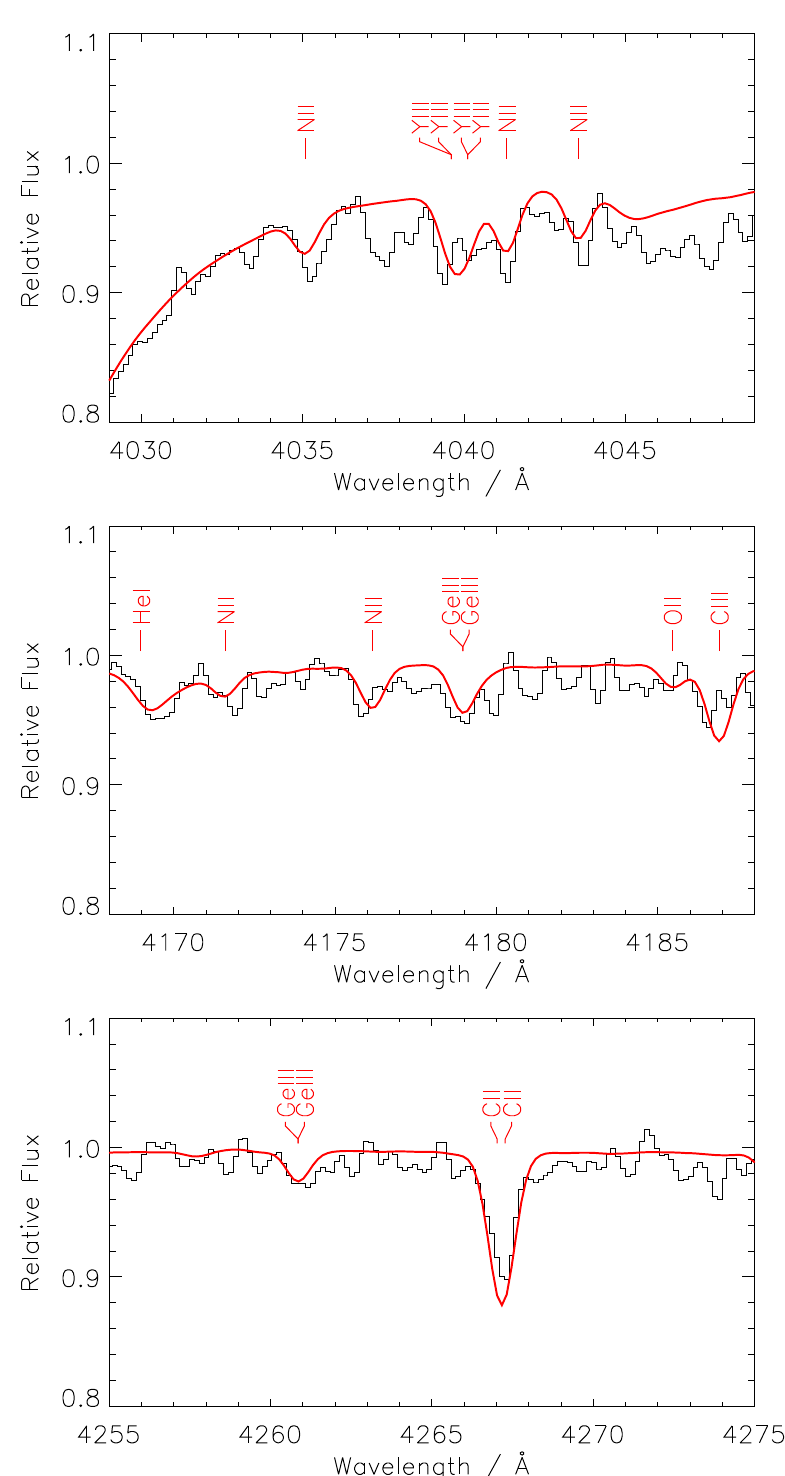,width=0.45\textwidth}
\epsfig{file=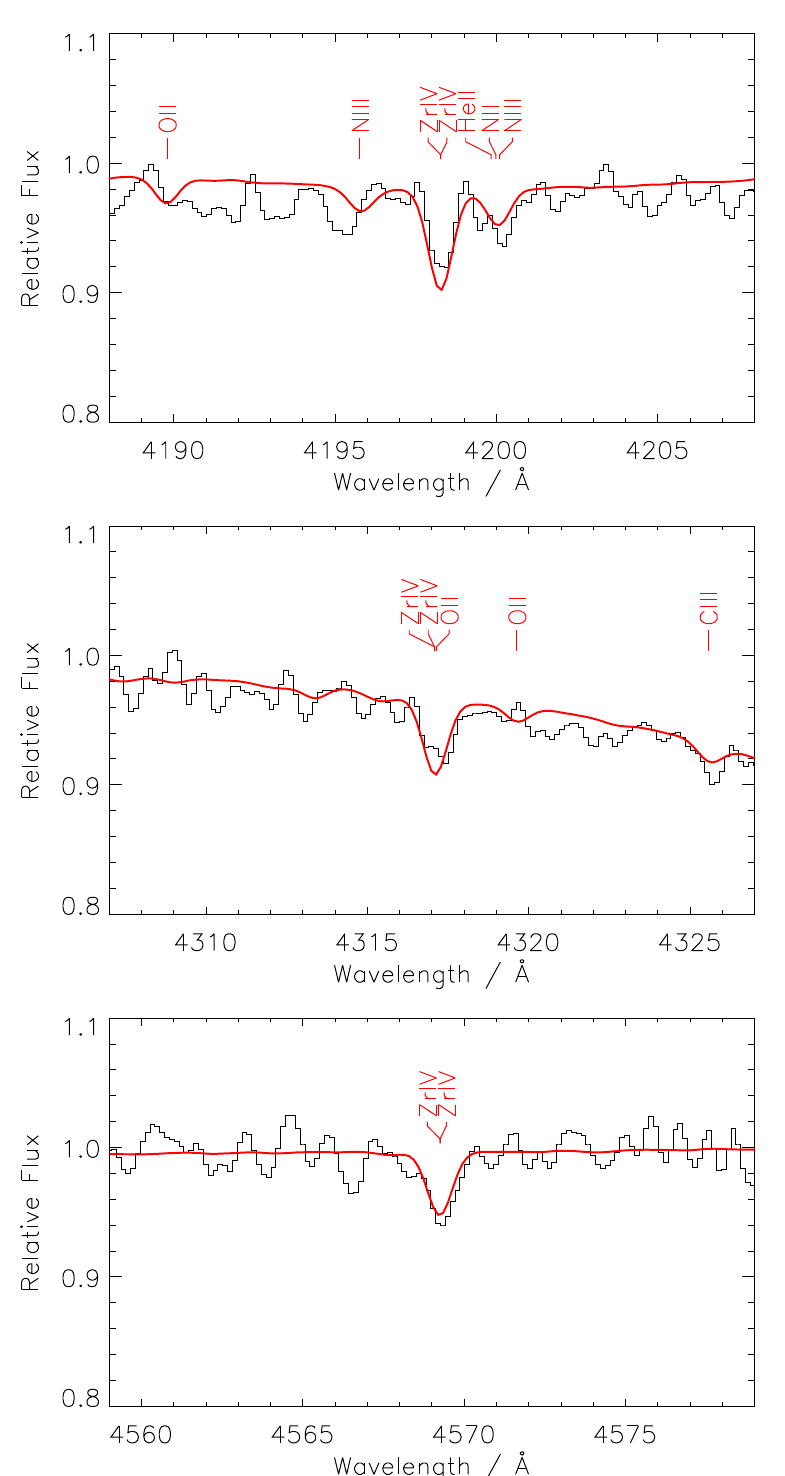,width=0.45\textwidth}
\caption{Selected regions of the MagE spectrum of \target\ (black histogram) and  model (red curve) as shown in Fig.\,\ref{f:fit_mage} showing the principal yttrium, germanium and zirconium lines. Other lines with predicted equivalent widths $>10$\,m\AA\ are labelled. }
\label{f:lines}
\end{figure*}

Photospheric parameters were measured from both the NOT/ALFOSC and MagE spectra using methods described by \citet{jeffery19}. 

The Armagh LTE radiative transfer package {\sc lte-codes} \citep{jeffery01b,behara06} includes a model atmosphere code {\sc sterne}, a formal solution code {\sc spectrum} and a general purpose fitting package {\sc sfit}\footnote{Local thermodynamic equilibrium was assumed throughout the analysis}. 
Effective temperature, surface gravity and surface helium and hydrogen abundances were obtained by finding the best fit spectrum in a grid of  models. To improve on any normalisation carried out during data reduction, a high-pass filter is applied during the fit, following the steps described by \citet{jeffery98a}. A broad enough filter is chosen so as not to degrade the gravity-sensitive line wings. The preliminary ALFOSC fit cross-checks that this has been applied correctly.  

For both spectra we use a grid\footnote{In this notation, 
the grid is defined by three parameters $[p_1$, $p_2$, $p_3]$ and  
three triplets $p_{\rm min} (\delta p) p_{\rm max}$ implying 
 $p \in p_{\rm min}, p_{\rm min}+\delta p, p_{\rm min}+2 \delta p, \ldots, p_{\rm max}$. Additional terms are added if the grid step $\delta p$ changes.}
covering a parameter space with  
\[
\begin{split} 
&[\teff/{\rm kK},\lgcs,\nHe ]  \\
& = [ 28(02)40 , 5.0(0.25)6.25,  0.1(0.1)0.3(0.2)0.90 ],
\end{split}
\]
sampled over the wavelength interval  $3400 - 7100$\AA , and which 
assumed an abundance distribution for elements heavier than helium based on a mean for hot subdwarfs compiled by \citet{pereira11}, but simplified for the {\sc sterne} input to 1/10 solar for $2 < Z < 26$ and solar for $Z \geq 26$ \citep{naslim20}. 
A microturbulent velocity $v_{\rm turb}= 5\kmsec$ was assumed for both the calculation of
line opacities in the model atmosphere calculation (which affects the temperature stratification of the models) and for the formal solution, which affects relative line strengths and widths. 
Whilst possibly overestimated for the current case, associated systematic errors are small compared with those arising from data quality \citep{jeffery19}. 

\begin{figure*}
       \epsfig{file=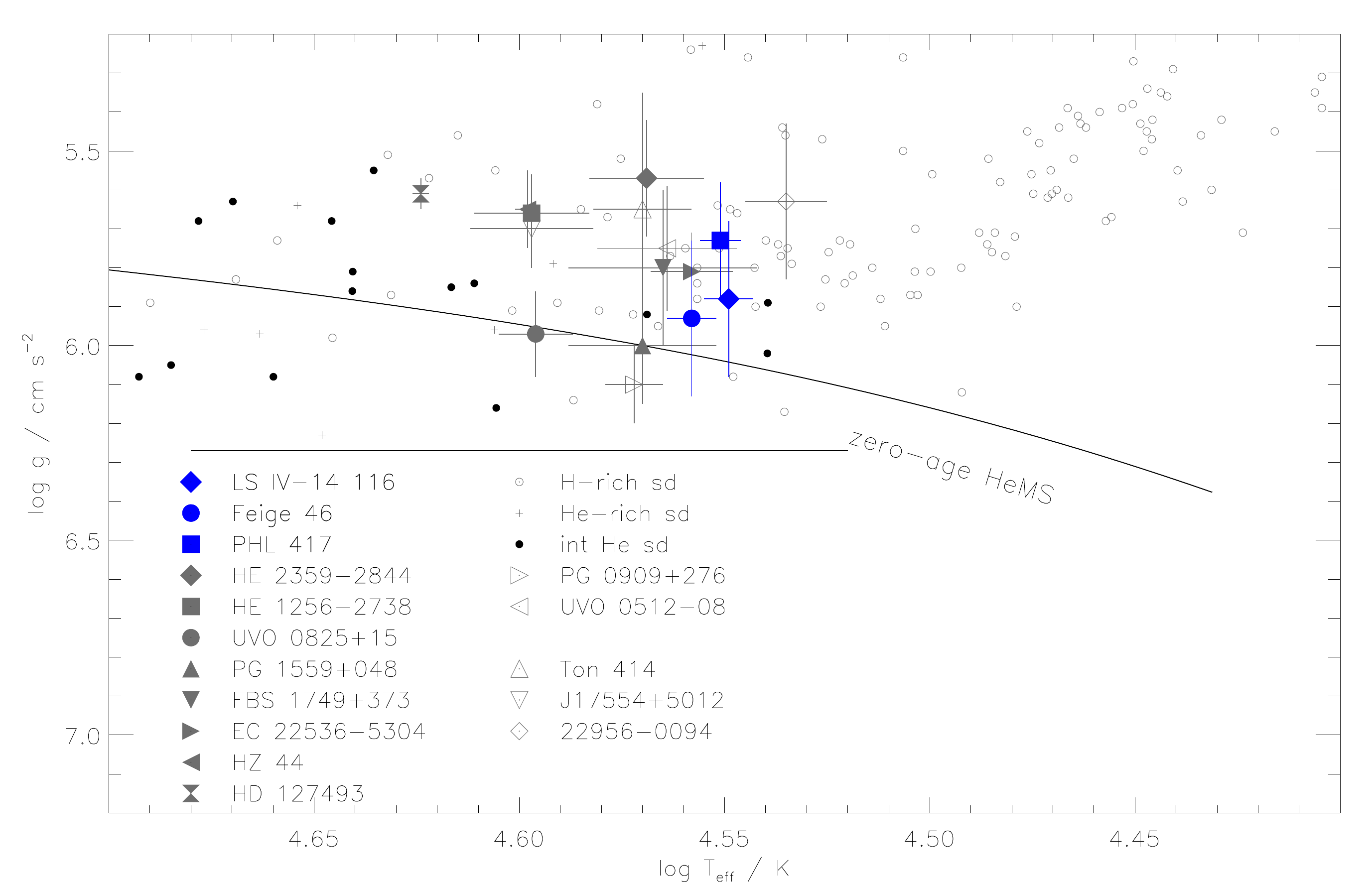,width=150mm}\\
       \epsfig{file=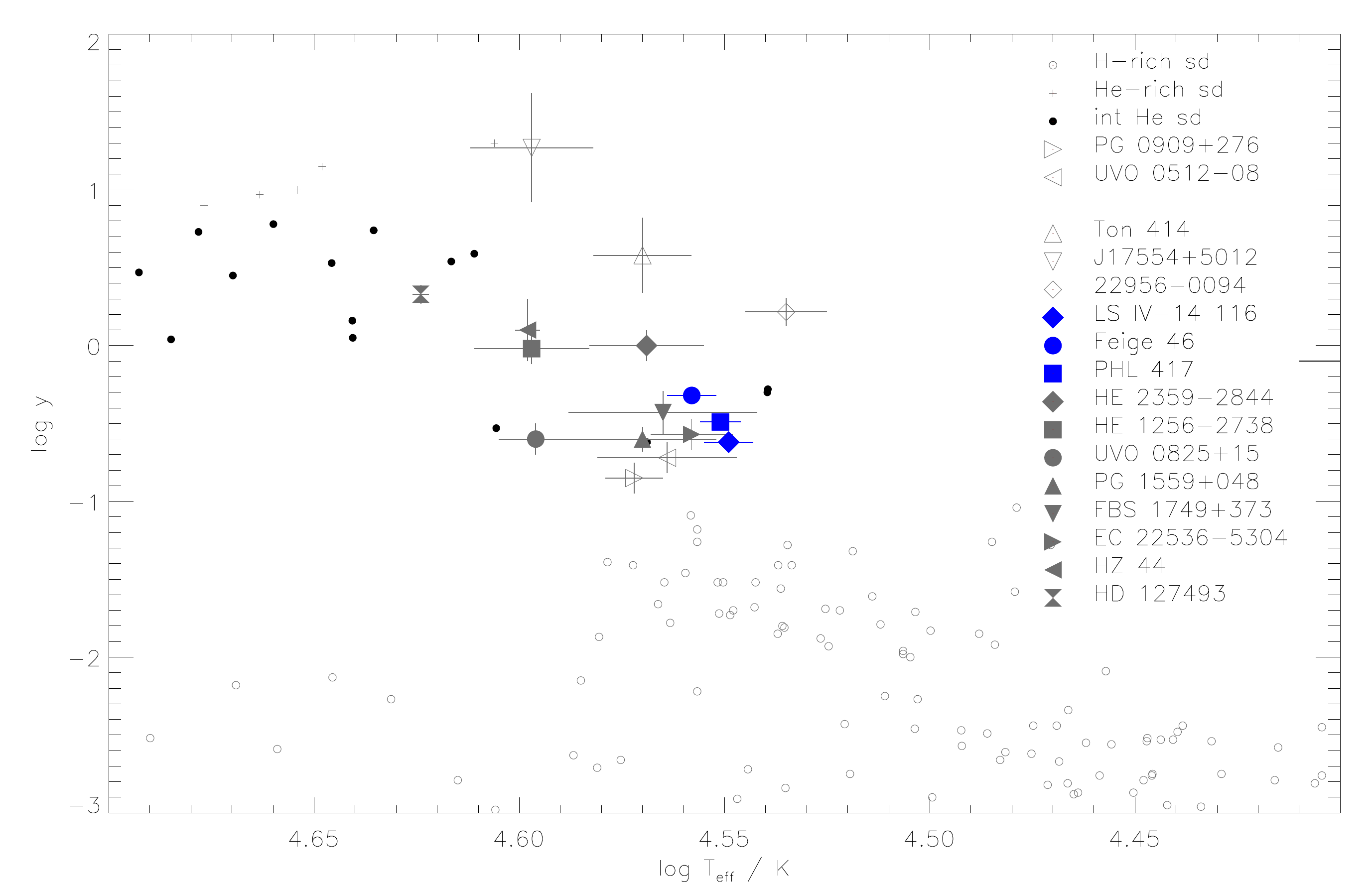,width=150mm}
        \caption{The distribution of V366-Aqr variables and other heavy-metal (large filled symbols), helium-rich and normal hot subdwarfs in surface gravity (top) and helium abudance (bottom) as a function of effective temperature. The large open symbols are intermediate helium-rich subdwarfs which do not show an excess of heavy metals. The solid line shows a representative position for the theoretical zero-age helium main-sequence (HeMS: $Z=0.02$). The data are taken from \citet{nemeth12,naslim13,randall15,jeffery17a,wild18,jeffery19,latour19b,dorsch19,naslim20} as well as this paper.  
} 
        \label{f:gt}
\end{figure*}

The solutions obtained in each case are given in Table\,\ref{t:atmos} (ALFOSC and MagE). 
The solution for the NOT/ALFOSC spectrum is shown in Fig.\,\ref{f:fit_alfosc}.
There are systematic differences associated with the different wavelength ranges adopted in the solutions, which mean that different hydrogen and helium lines contribute, in particular, to the surface gravity measurement and hence in turn to the effective temperature and surface helium, since these quantities are not independent in the solutions. 
We  adopt the mean of the ALFOSC and MagE measurements for comparison with other hot subwarfs, giving $T_{\rm eff}=35.64\pm0.28$\,kK, $\log g / {\rm cm\,s^{-2}}=5.73\pm0.01$ and $\log y = -0.49\pm0.02$ (Fig.\ref{f:gt}). Errors are a combination of statistical errors in the fit and the difference between ALFOSC and MagE from Table\,\ref{t:atmos}.

These measurements are subject to a range of systematic effects, including the adoption of the LTE approximation, the metallicity and microturbulent velocity adopted in the models, the resolution and extent of the model grid, and the methods used to match the overall continuum level. 
Experiments with different model grids, including one based on zero-metal non-LTE model atmospheres \citep{nemeth12}, one with a solar distribution of metals,  and several with different distributions of grid points, yield results with ranges $\Delta\Teff \approx 1.87$kK, $\Delta \log g \approx 0.33$, and $\Delta \log y \approx 0.04$. 
Given the quality of the observational data, the above solution is taken to be indicative and we conservatively adopt systematic errors  $\delta\Teff \approx 0.75$kK, $\delta \log g \approx 0.15$, and $\delta \log y \approx 0.04$, which we combine quadratically with the statistical errors to give total errors $\Delta\Teff \approx 0.8$kK, $\Delta \log g \approx 0.15$, and $\Delta \log y \approx 0.05$.

\begin{table}
\caption[Line abundances]
   {Line abundances for PHL\,417 derived from the MagE spectrum showing wavelength ($\lambda$), adopted oscillator strengths ($\log gf$), equivalent widths ($W_{\lambda}$) and elemental abundances ($\log \epsilon$). The mean error on the equivalent widths is $\pm15$m\AA\ and is propagated to obtain the error on each line abundance. $d \epsilon/d W_{\lambda}$ depends where on the curve of growth $W_{\lambda}$ lies and differs from line to line by large amounts.   }
\label{t:lines}
\small
\begin{center}
\begin{tabular}{l rrcr }
\hline
Ion & $\lambda/$\AA & $\log gf^1$ & $W_{\lambda}/$m\AA & $\log \epsilon$ \\
\hline
C{\sc ii}  &  4744.77  &$-$0.600 &    27 & 7.93$\pm$0.31 \\
C{\sc ii}  &  4267.27  &   0.734 &    95 & 8.17$\pm$0.14 \\
C{\sc iii} &  4186.90  &   0.924 &    39 & 7.96$\pm$0.29 \\
C{\sc iii} &  4325.56  &   0.090 &    22 & 8.23$\pm$0.45 \\
N{\sc ii}  &  3995.00  &   0.225 &    52 & 8.03$\pm$0.26 \\
N{\sc ii}  &  5005.15  &   0.612 &    66 & 8.39$\pm$0.25 \\
N{\sc iii} &  4640.64  &   0.140 &    35 & 7.79$\pm$0.34 \\
O{\sc ii}  &  4075.86  &   0.693 &    37 & 7.83$\pm$0.31 \\
O{\sc ii}  &  4072.15  &   0.552 &    28 & 7.77$\pm$0.36 \\
Ne{\sc ii} &  3875.28  &$-$0.080 &    39 & 7.67$\pm$0.24 \\
Si{\sc iv} &  4088.85  &   0.199 &    13 & 5.57$\pm$0.67 \\
P{\sc iii} &  4222.20  &   0.210 &    22 & 5.99$\pm$0.36 \\
S{\sc iii} &  4253.59  &   0.400 &    15 & 6.13$\pm$0.50 \\
S{\sc iii} &  4284.98  &   0.110 &    16 & 6.44$\pm$0.47 \\
Ge{\sc iii}&  4178.96  &   0.341 &    31 & 5.91$\pm$0.30 \\
Ge{\sc iii}&  4260.85  &   0.108 &    25 & 6.02$\pm$0.35 \\
Y{\sc iii} &  4039.60  &   1.005 &    48 & 6.48$\pm$0.22 \\
Zr{\sc iv} &  4198.27  &   0.323 &    56 & 6.10$\pm$0.20 \\
Zr{\sc iv} &  4317.08  &   0.069 &    47 & 6.24$\pm$0.22 \\
Zr{\sc iv} &  4569.25  &   1.127 &    59 & 6.63$\pm$0.19 \\
Zr{\sc iv} &  4137.44  &$-$0.625 &    33 & 6.68$\pm$0.28 \\
\hline
\end{tabular}\\
\parbox{70mm}{
$^1$; 
C{\sc ii}: \citet{yan87}, 
C{\sc iii}: \citet{hib76,har70}, 
N{\sc ii}: \citet{bec89}, 
N{\sc iii}: \citet{but84}, 
O{\sc ii}: \citet{Bec88},
Ne{\sc ii}: \citet{wie66},
Si{\sc iv}: \citet{bec90}, 
P{\sc iii}: \citet{wie66}, 
S{\sc iii}: \citet{wie69},
Ge{\sc iii}: \citet{naslim11},
Y{\sc iii}: \citet{naslim11},
Zr{\sc iv}: \citet{naslim11}. }
\end{center}
\end{table}

\begin{table*}
\caption[Atmospheric abundances]
   {Atmospheric abundances of \target\ and other V366\,Aqr variables given as $\log \epsilon$, normalised to $\log \Sigma \mu \epsilon = 12.15$. Errors are given in parentheses. The second line, marked $n$, shows the number of lines contributing to each measurement for \target. 
   }
   \label{t:abunds}
\setlength{\tabcolsep}{3pt}
\begin{tabular}{@{\extracolsep{0pt}}p{20mm} ll lll lll lll ll}
\hline
Star 	& H & He & C & N & O	& Ne & Si & P &   S & Ge & Y   & Zr & Reference \\
\hline        
PHL\,417 & 11.77(3) & 11.31(3) & 8.0(2) & 8.1(3) & 7.8(2) & 7.7(2) & 5.6(7) & 6.0(4) & 6.3(4) & 6.0(2) & 6.5(2) & 6.4(3) & This work \\
$n$ &  &  & 4 & 3 & 2 & 1 & 1 & 1 & 2 & 2 & 1 & 4 \\[2mm]
\lsiv & 11.83(7) & 11.15(5) & 8.0(2) & 8.0(2) & 7.6(2) & $<7.6$ & 6.3(1) & & & 6.3(1) & 6.2(1) & 6.5(2) & \citet{naslim11} \\
Feige\,46 & 11.71(2) & 11.35(4) & 8.6(3) & 8.1(1) &  7.5(2) & --  & 5.6(1) & &  $<5.9(3)$& 5.8(6) & 6.5(4) & 6.6(1) & \citet{latour19b}\\[2mm]
Sun & 12.00 & 10.93 & 8.4 & 7.8 & 8.7 & 7.9 & 7.5 & 5.4  & 7.1 & 3.7 & 2.0 & 2.6 & \citet{asplund09} \\
\hline
\end{tabular}\\
\end{table*}


\section{Abundances}

We computed a model atmosphere close to the MagE solution and proceeded to derive a formal solution having abundances corresponding to the mixture adopted above. 
Given the pulsation properties we experimented with increasing the zirconium abundance and immediately recognised three strong zirconium lines in the MagE spectrum. 
The resolution and signal-to-noise of the latter are not ideal for abundance analysis. Nevertheless, the presence (or absence) of certain lines can be used to obtain some idea of the surface composition (Fig.\,\ref{f:fit_mage}). 
Predictably present in the wavelength range 3850\,--\,5100\,\AA\ are lines of carbon (C{\sc ii,iii}), nitrogen (N{\sc ii,iii}), oxygen (O{\sc ii}), neon (Ne{\sc ii}), silicon (Si{\sc iv}), sulphur (S{\sc iii}) and phosphorous (P{\sc iii}).  Also and remarkably present are zirconium (Zr{\sc iv}), germanium (Ge{\sc iii}) and probably yttrium (Y{\sc iii}) (Fig.\,\ref{f:lines}). 
There was no evidence for lead (Pb{\sc iv}).  
Equivalent widths were measured for the strongest lines for all of these species; the threshold is about 12 -- 15 m\AA, and this also represents the typical error on all of the measured equivalent widths.
A model atmosphere was computed corresponding to the final parameters adopted in Table\,\ref{t:atmos}. 
Table~\ref{t:lines} shows abundances derived from the measured equivalent widths assuming a micro-turbulent velocity $v_{\rm turb}=5$\kmsec. 
Abundances obtained for C{\sc ii} and C{\sc iii} are similar, indicating that equilibrium is satisfied for this pair of ions.
A formal solution based on these abundances is shown in Fig.\,\ref{f:fit_mage}.
The results are summarised in Table~\ref{t:abunds}, where it is emphasised that these are based on a few strong lines and higher precision measurements are desirable.
For comparison, Table~\ref{t:abunds} also shows the solar abundances for the same species; 
note that atmospheric carbon and oxygen are depleted in \target\  with respect to solar.
Statistical errors cited for $\log \epsilon_i$ are derived from equivalent width measurement errors and/or the standard deviation around the mean of multiple line measurements. 
Systematic errors in $\log \epsilon_i$  propagated from total errors $\delta \log g / {\rm cm\,s^{-2}} \approx 0.15$, $\delta \nHe\approx 0.05$ and $\delta v_{\rm turb} \simle 5$\kmsec are all $\simle0.05$. 
The total error  $\delta \teff \approx 0.8$kK leads to systematic errors $|\delta \log \epsilon|$ ranging from $<0.06$ (N, C, Ne ) up to $\approx 0.36$ (P, Ge) with the remainder ranging from 0.08 to 0.23 (Zr, Y, O, Si, S). 
The systematic abundance errors are all smaller than the statistical errors.

The preceding analysis shows that  \target\ belongs to the group of intermediate helium subdwarfs (iHe-sds) as defined by \citet{naslim10}. 
These objects distinguish themselves from the main population of sdB stars, which are clearly associated with the EHB, since they produce a narrow band in the $g-\teff$ diagram (Fig.\,\ref{f:gt}). 
They are also significantly different from the rather large population of very helium-rich sdO and sdOB stars, which are found in the $g-\teff$ diagram mostly between $\sim40$ and 50\,kK without any clustering in $g$. 

Among a number of proposed solutions, one possible explanation for these objects is that they are evolving towards the helium main sequence, or zero-age EHB, and that their low atmospheric H abundance is due to gravitational stratification not having had the time to act since their formation {\bf \citep{naslim10,naslim11}}.
The intermediate helium stars may represent a transition between these objects, at least for those objects that are found close to the zero-age EHB or He-MS.
This is exactly where we find the three V366-Aqr pulsators.

\target\ is clearly related to \lsiv\ and Feige\,46 as it has similar spectroscopic parameters and surface composition (as far as the data quality permits).  The so-called heavy-metal stars all lie close to the EHB/HeMS junction in the $g-\teff$ diagram. 
They fall into three distinct groups: 
a) the V366\,Aquarids, which are rich in germanium, yttrium and zirconium (solid blue symbols in Fig.\,\ref{f:gt}),  
b) a group of about eight stars, including UVO\,0825+15, which show very significant overabundances of lead and other elements with $Z\geq20$, including in some instances, zirconium (solid grey), 
and c) a number of other intermediate helium stars with very high overabundances of iron-group elements (open grey).

\begin{figure}
  \includegraphics[width=1.0\linewidth]{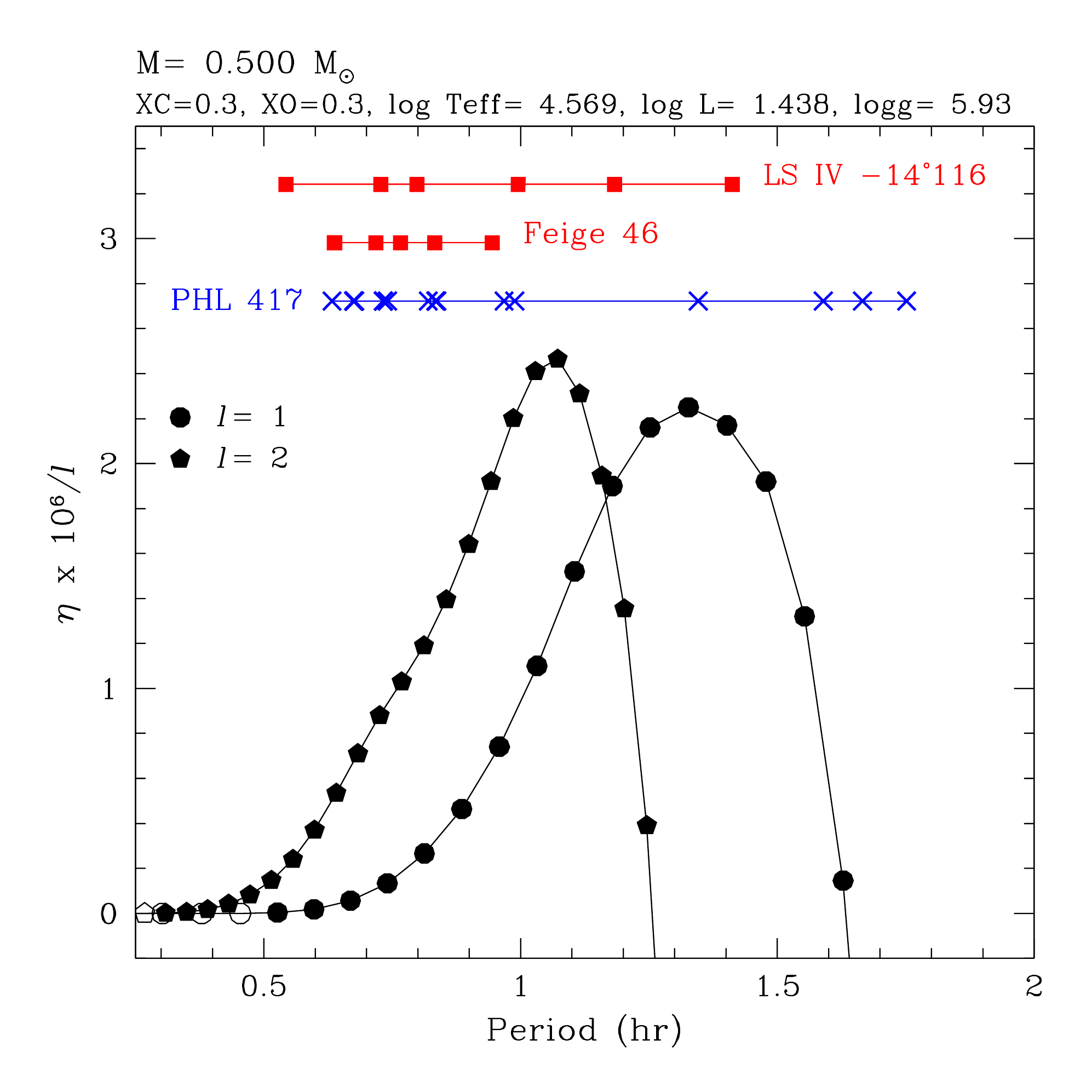}
\caption{Growth rates $\eta$ for predicted  modes for a 0.50\Msolar\ helium main-sequence star enriched by 30\% carbon and 30\% oxygen (by mass) are shown as a function of period.
Stable modes are shown as open symbols, unstable models as solid black symbols.
\ellone\ and \elltwo\ modes are shown as circles and pentagons respectively. 
The observed pulsation periods for all three Aquarids are shown and labelled.}
\label{f:pulse}
\end{figure}

\begin{figure}
  \includegraphics[width=1.0\linewidth]{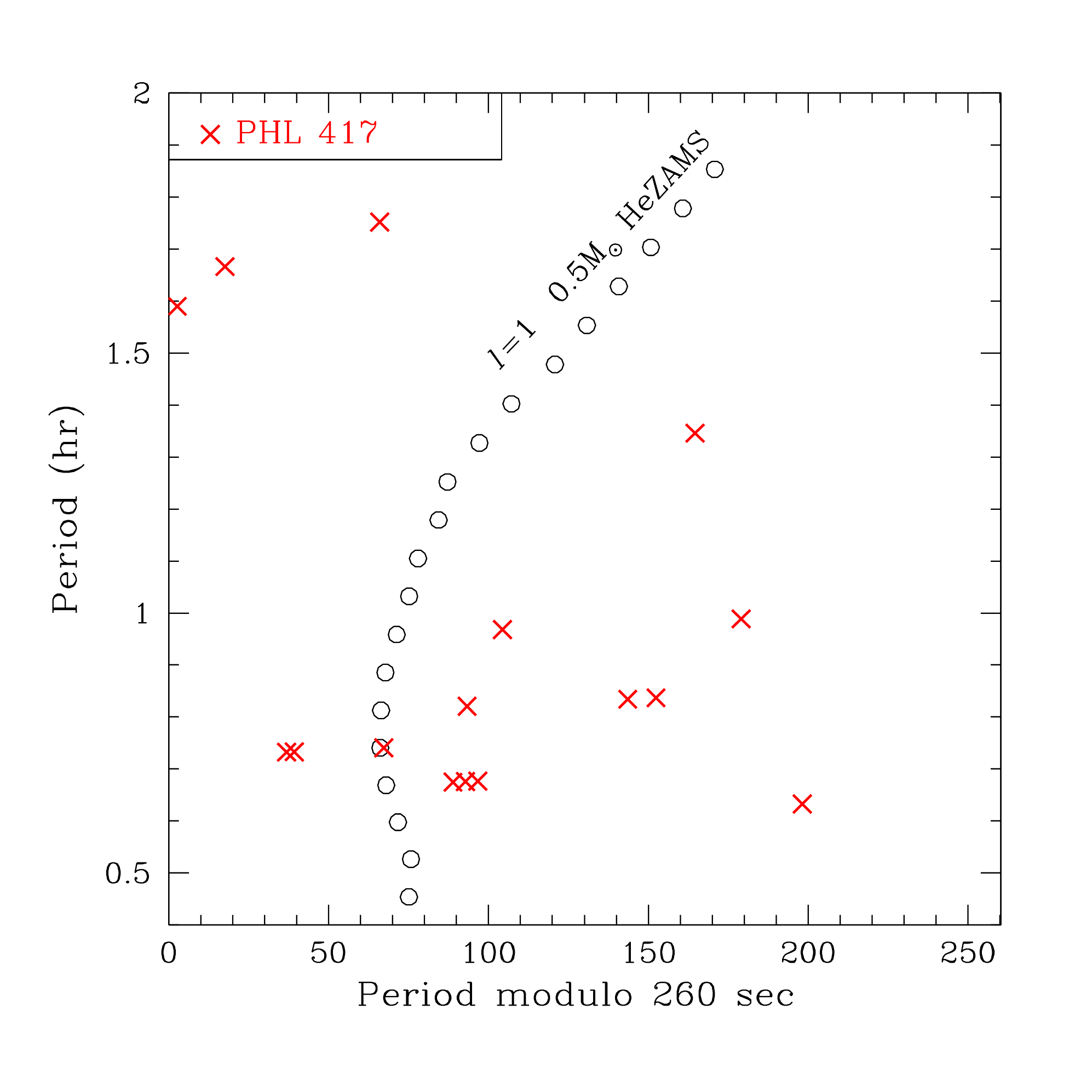}
\caption{The \'echelle diagram for \ellone\ modes for the model shown in Fig.~\ref{f:pulse}. Modes observed in \target\ are plotted as red $\times$ (grey in print). }
\label{f:echelle}
\end{figure}

\section{Pulsations}

The discovery of pulsations in \lsiv\ \citep{ahmad05,green11} presented a problem since, with $\teff\approx35\,500$\,K, the star is too hot for g-modes to be driven by Fe\,+\,Ni bump opacities. 
\citet{bertolami11} and \citet{battich18} have proposed that a nuclear instability in the inward-burning shell of a pre-EHB star can excite pulsations via the $\epsilon$-mechanism. 
Again, the difficulty is to drive pulsations in stars with precisely the  \teff\ and $g$ ranges of \lsiv\ and, also, Feige\,46 \citep{latour19a,latour19b}. 
The pulsations in \target\ span the same period range and also have similar amplitudes (0.3 -- 3.3 ppt) to Feige\,46 (0.9 -- 2.9 ppt) and \lsiv (1.0 - 2.7 ppt) \citep{green11,latour19a}.
By comparing thermal and dynamical timescales of models in the correct \teff\ and $g$ ranges, \citet{saio19} proposed that the $\kappa$-mechanism would work if there were opacity bumps around $10^6$\,K. 
By constructing a class of helium core-burning models in which the abundances of carbon and oxygen are substantially enhanced at these temperatures, \citet{saio19} found that g-mode pulsations could be excited in exactly the period ranges observed in \lsiv\ and Feige\,46.

Following discovery of pulsations in \target\ and using the same methods as described in \citet{saio19}, a new model was constructed having approximately the observed surface properties of \target, an assumed mass of 0.50 \Msolar, and an interior composition of 40\% helium, 30\% carbon and 30\% oxygen.  
Such a structure might arise, for example, if the subdwarf evolved from a relatively massive zero-age helium main-sequence star, after a very rapid mass-loss event due to interaction with a companion in a binary system \citet{saio19}.  
Other metals have scaled-solar ratios with $Z=0.02$. 
Considering both dipole (\ellone) and quadrupole (\elltwo) modes, g-modes are excited across the entire period range observed in \target\ (Fig.\,\ref{f:pulse}).   

All three V366\,Aqr variables are pure g-mode pulsators with nothing detected in the p-mode regime, even at the photometric sensitivity of {\it K2}. 
Note that p-modes in these stars are predicted to have periods of less than about 300\,s, which would be easily detectable in {\em Kepler} short-cadence photometry, where the Nyquist limit is around 118\,s. 
The $\kappa$-mechanism excitation in an ionization zone is most effective if the period of pulsation is comparable with the thermal timescale there \citep[e.g.][]{cox74}. 
However, in the models of V366-Aqr variables, p-mode oscillations cannot be excited by the $\kappa$-mechanism in helium or C/O ionization zones, because the thermal timescale is too short ($\sim0.1$\,s) in the helium ionization zone or too long ($\sim10^3$ to $10^5$\,s) in the C/O ionization zone.

A feature of the FT presented in Fig.\,\ref{fig:kepft} is the relatively high density of g-modes and the absence of any asymptotic sequence. 
Figure\,\ref{f:echelle} shows an \'echelle diagram for \target\ with \ellone\ g-modes of a 0.5\,\Msolar\ model. 
The observed g-mode density is higher than expected for \ellone\ and 2 g-modes alone. 
Relatively little else may be concluded except that the observed triplet is most likely a dipole mode.

\section{Conclusions}

{\em K2} photometry of \target\ has revealed a relatively rich pulsation spectrum in the g-mode region of pulsations in hot subdwarf stars.
Follow-up ground-based spectroscopy demonstrates physical parameters and abundances that have allowed us to identify this object as the third member of the rare group of pulsating heavy-metal stars, the V366\,Aquarii pulsators.

The nomenclature is an important aspect of astronomy. We have chosen to use the variable star designation, V366\,Aqr, to label the new class of pulsators, as is traditional in the variable star community. This distinguishes the new pulsators from the hitherto recognised classes of helium-poor sdB pulsators, the short period V361\,Hya and long-period V1093\,Her classes. The new class is characterised by having a spectral composition similar to the prototype, \lsiv, and pulsations with periods between 0.5 and 2\,h. An alternative would have been to follow the pattern preferred by some of adding a V to the spectroscopic designation. But spectroscopic designations can be exceedingly complex, and often change as better observations provides a more detailed picture. For the intermediate-helium subdwarfs, the designation iHe-sdB has been used by some authors, although it should more precisely be iHe-sdOBZ, in order to recognise the presence of both He{\sc i} and He{\sc ii} as well as heavy metals in their optical spectra. We prefer to avoid this path. The pattern used for pulsating CO white dwarfs is simpler and less ambiguous. The star PG\,1159$-$035 is the prototype of this spectroscopic class, and  stars with similar spectra are referred to as PG\,1159 stars. It was also the first star of its class to be discovered to pulsate, and given the variable star designation GW\,Vir. The subset of PG\,1159 stars that are pulsating are therefore referred to as GW-Vir pulsators (although some authors still prefer to designate them as DOV or PNNVs). Following this pattern, we suggest that the heavy-metal stars be referred to as \lsiv\ stars, and pulsating members of this spectroscopic class as V366-Aqr stars.

The two K2 runs demonstrate that the pulsations are stable when comparing frequencies in observations separated by 1.5 years, with an upper limit on $\dot{P}$ of $1.1\cdot10^{-9}$ s/s from the highest amplitude mode.
This suggests that the interior of the star cannot be rapidly changing.
The recent theoretical predictions that pulsations in heavy-metal stars are driven during an active core-helium flash event, and therefore should change with rates as high as $\dot{P}$\,$\approx$\,$10^{-7}$\,--\,$10^{-4}$\,s/s is not consistent with the upper limits on period change we establish for PHL\,417.

\section*{Acknowledgments}

RH\O, ASB, MDR gratefully acknowledge financial support from the Polish National Science Center under projects No.\,UMO-2017/26/E/ST9/00703 and UMO-2017/25/B/ST9/02218.
JV acknowledge financial support from FONDECYT grant number 3160504. JJH acknowledges support by NASA K2 Cycle 6 Grant 80NSSC19K0162.

This paper includes data collected by the K2 mission. Funding for the K2 mission is provided by the NASA Science Mission directorate.

The spectroscopic observations used in this work were collected at the Nordic Optical Telescope at the Observatorio del Roque de los Muchachos on La Palma, and at the Magellan Telescope at Las Campanas Observatory, Chile.

\section*{Data Availability}
The photometric data underlying this article are in the public domain and can be accessed from the Barbara A. Mikulski Archive for Space Telescopes at {\tt mast.stsci.edu/portal/Mashup/Clients/Mast/Portal.html}  with either of the identifiers PHL\,417 or EPIC\,246373305. 
The spectroscopic data underlying this article will be shared on reasonable request to the corresponding author.

\bibliographystyle{mn2e}
\bibliography{sdbrefs,atomic}
\label{lastpage}
\end{document}